\documentclass[11pt]{article}

\usepackage{amsmath,amsfonts,xypic}

\usepackage{amssymb}

\usepackage{amscd}

\usepackage{graphicx}

\usepackage{undertilde}

\usepackage{multirow}

\advance \textheight by \topskip

\makeatletter

\@addtoreset{equation}{section}

\makeatother

\def\ZZ{{\mathbb Z}}

\newcommand{\beq}{\begin{equation}}

\newcommand{\eeq}{\end{equation}}

\newcommand{\beqa}{\begin{eqnarray}}

\newcommand{\eeqa}{\end{eqnarray}}

\newcommand{\noi}{\noindent}

\newcommand{\e}{\varepsilon}

\newcommand{\g}{{\mathfrak g}}

\def\lb{\left\{\left|}

\def\gr{\text{gr}}

\def\rb{\right|\right\}_{_{\hskip -.05 truecm N}}}

\def\>{\rangle}

\def\<{\langle}

\usepackage{color}
\newcommand{\nc}{\newcommand}
\nc{\rr}{\color{red}}
\nc{\bl}{\color{blue}}
\nc{\gre}{\color{green}}

\begin{document}

\title{
{\bf  Parafermions for higher order extensions of the Poincar\'e
algebra and their associated superspace}  }
\author{
{\sf  R. Campoamor-Stursberg}\thanks{e-mail:
rutwig@pdi.ucm.es}$\,\,$${}^{a}$ and {\sf  M. Rausch de
Traubenberg}\thanks{e-mail: Michel.Rausch@IReS.in2p3.fr
}$\,\,$${}^{b}$
\\
{\small ${}^{a}${\it I.M.I-U.C.M,
Plaza de Ciencias 3, E-28040 Madrid, Spain}}  \\
{\small ${}^{b}${\it IPHC-DRS, UdS, CNRS, IN2P3; 23  rue du Loess,
}}\\
{\small {\it  Strasbourg, 67037 Cedex, France}} }
\date{\today}
\maketitle \vskip-1.5cm

\begin{abstract}
Parafermions of order two and three are shown to be the
fundamental tool to construct superspaces related to cubic and
quartic extensions of the Poincar\'e algebra. The corresponding
superfields are constructed, and some of their main properties
analyzed in detail. In this context, the existence problem of
operators acting like covariant derivatives is analyzed, and the
associated operators are explicitly constructed.
\end{abstract}

\vspace{2truecm}

\section{Introduction}

The question whether symmetric and antisymmetric functions are the
only physically acceptable types for eigenfunctions can be traced
back to the beginning of Quantum Mechanics, and was generally
answered in the affirmative until the discovery of new particles
and resonances forced to reconsider the possibility of alternative
symmetry types.

The first developments on intermediate statistics were developed
in 1940, at the same time of Pauli's theorem \cite{GeN}. In this
work, the combinatorial method of Bose was used to infer an
expression for the average number of particles in a set of states
that is independent on the maximal number of particles that occupy
a given state. The ansatz was mainly based on the first-quantised
formulation of statistics, and therefore this approach was not
entirely satisfactory from the quantum field theoretic point of
view. In 1953, Green introduced what is nowadays known as
paraquantisation \cite{green,para}, leading to two families of
generalised statistics, each one containing one of the classical
Bose and Fermi statistics types. This pioneering work was later
developed and refined by Greenberg and Messiah \cite{GrB,GM}, who
settled under which conditions parastatistics do not contradict
established experimental facts.

The discovery of the $\Omega^{-}$-hadron in 1964 not only supposed
one of the first successes of the symmetry approach to elementary
particle classification, to be further worked out in subsequent
years (finally leading to the Standard Model), but also pointed
out some difficulties that were not entirely understood and
explained until the overflow of particle discoveries in the 50's
and 60's. The quark content of $\Omega^{-}$ following from the
flavour classification of hadrons apparently meant an evidence to
the negative for the Pauli exclusion principle, which reopened the
discussion on its range of validity and its exact interpretation
with respect to the underlying statistics. This fact, joined to
other minor incompatibilities observed earlier, showed that some
fundamental characteristics were still to be discovered, and
suggested that the newly introduced quarks obeyed not the usual
Fermi statistics, but some type of intermediate statistics.
Actually this assumption on the para-particle character of quarks
provided an alternative solution to the introduction of a
threefold degree of freedom for quarks (the colour quantum number)
that gave birth to Quantum Chromodynamics (QCD). However, since
parastatistics was not amenable to gauging, the idea of
paraquantisation as a description of a fundamental symmetry for
the quarks was abandoned to the benefit of QCD or more generally of gauge
theory.

Subsequently, the gauge principle becomes central in the
description of fundamental interactions. This principle together
with a series of no-go theorems \cite{no-go} lead to the dominant
framework for a description of physics beyond the Standard Model,
namely to  supersymmetry or  supergravity. Supersymmetry (resp.
gauge theories) are based on Lie superalgebras (resp. Lie
algebras) which are binary algebras.  In spite of the great
success of gauge theories and supergravity,
 one may wonder whether or not some different algebraic structures
could play a role in physics, and in particular higher order
algebras. This question was rather academic and binary algebras
(Lie superalgebras) were dominating the description of the
symmetries in particles physics until recently. Indeed, it was
realised that higher order algebras could play some roles in
physics. For instance,
 a  ternary algebra defined by a fully
antisymmetric product appears in the description of  multiple
M2-branes \cite{bl}. Similarly, higher order extensions of the
Poincar\'e algebra were defined without contradicting the no-go
theorems \cite{flie} and implemented into the Quantum Field Theory
(short QFT) frame. Then, despite many efforts, the construction of
an adapted ``superspace'' associated to these latter higher order
extensions was not known. The purpose of this paper is to show
that parafermions are the basing building block for the
construction of a superspace associated to the higher order
extensions of the Poincar\'e algebra considered in \cite{flie}.
This means that parafermions reappear for the description of
symmetries in physics, but in a different context compared to its
historical consideration.

\smallskip

In a series of papers, $F-$ary extensions ($F>2$) of Lie superalgebras,
called Lie algebras of order $F$, were introduced and analysed
\cite{flie,gr,hopf,color}. It was then rapidly realised that these
new algebraic structures could be used  to define higher order
extensions of the Poincar\'e algebra. Among various possibilities,
a specific cubic extension of the Poincar\'e algebra in any
space-time dimensions was intensively studied in
\cite{cubic1,cubic2,p-form} and implemented into QFT. However,
this program to investigate new types of non-trivial extensions of
the Poincar\'e algebra partially failed. Indeed, at present, only
invariant free Lagrangians have been constructed. It was even
proven that, in four space-time dimensions, for specific types of
multiplets, no interaction terms were possible \cite{cubic2}. In order to
construct invariant interacting Lagrangians, one of the most
promising alternatives  would be to identify some adapted
``$F-$ary superspace'', where the higher order symmetries would be
realised by means of differential operators. Such a construction
has indeed been considered by several authors in one or two
space-time dimensions, where the situation is somewhat exceptional
(the Lorentz algebra being either trivial or abelian)
\cite{super}, and a nice geometrical interpretation was given in
\cite{az}. It seemed however that the success of such models were
deeply related to the low dimension, and as soon as the space-time
dimension is higher than three, no corresponding superspace has
been identified. This obstruction is certainly due, at least in
part, to the type of algebras considered, which are ternary (and
in general $F-$ary) instead of binary.

Among the main difficulties arising in non-binary approaches, we
observe that it is not possible to order a given monomial in a
definite order, implying that finite dimensional representations
are automatically non-faithful \cite{flie,hopf}. Despite these
difficulties, it was recently realised that Lie algebras of order
$F$ share some similarities with Lie superalgebras. A formal study
of Lie algebras of order three gives rise to two interesting
results. The first important result in this direction is the
construction of groups associated to Lie algebras of order three.
Furthermore, the parameters of the transformations have been
identified, and correspond to the natural cubic extension of the
Grassmann algebra, called the three-exterior algebra (see
\eqref{roby} below) introduced by Roby \cite{roby}. These
similarities enable the construction of matrix representations of
groups associated to Lie algebras of order three, in terms of
matrices, the entries of which belong to the three-exterior
algebra \cite{hopf}, in straight analogy with Lie supergroups.

Since the parameters of the transformations generate the
three-exterior algebra, it is natural to postulate that these
variables  generate also the superspace corresponding to the
ternary extensions of the Poincar\'e algebra we are considering.
Imposing that a differential realisation of the algebra is
obtained by means of the new variables and some associated
differential operator,  automatically leads to a parafermionic
algebra \cite{green,para}. It is very interesting to notice that
two different structures, which have {\it a priori} no relation,
can be unified by this ansatz. The question whether these two
structures (parafermions and Lie algebras of order $F$) have some
hidden relations arises at once.

The contents of this paper is the following. In section 2. the
mean features of Lie algebras of order $F$ ($F>1$) are recalled
together with some emphasis on the cubic and quartic extensions of
the Poincar\'e algebra relevant in the sequel. In section 3. it is
shown that  parafermions (of order two for ternary extensions and
of order three for quaternary extensions) are the fundamental
objects to define  a  superspace associated to the cubic/quartic
extension of the Poincar\'e algebra considered. It is also shown,
studying quartic extensions in any space-time dimensions that the
case $D=1+9$ is very special. In this particular case a quaternary
superspace can be constructed using {\it usual fermions}.
Moreover, it is interesting to notice that this quartic extension
presents some analogy with type IIA supersymmetry. Section 4 is
devoted to the study of two types of superfields, as well as the
construction of certain operators which can be interpreted as a
covariant derivative. It turns out that the implications of these
operators for cubic and quartic extensions differ in some
fundamental aspects. Some conclusions on these
constructions are drawn in section 5.\\

Finally, we mention for completeness that there has been also some
revival of interest for parafermions and parabosons, in a rather
different context. In \cite{tern} it was realised that
paraquantisation is related to Lie superalgebras, and in
\cite{parag} parafermions were the basic building block to
construct some parafermionic extensions of the Poincar\'e algebra
in the context of para-superalgebras. In \cite{ply}, 
it was shown that some purely parabosonic (parfermionic) systems
are described by a hidden nonlinear (polynomial)  supersymmetric
quantum mechanics.

\section{Lie algebras of order three}
In this section, we recall the basic properties of Lie algebras of
order $F>2$. We also recall the main features of  cubic and
quartic extensions of the Poincar\'e algebra that will be relevant
in the sequel. Higher order algebraic structures, called Lie
algebras of order $F$ and  generalising Lie (super)algebras were
introduced in \cite{flie}.  Complex and real Lie algebras of any
order $F>2$ may be defined. In this paper we study more precisely
elementary  real Lie algebras of order three and four. An
elementary (real) Lie algebra of order $F$ is given by $\g= \g_0
\oplus \g_1 = \left<X_i,  \ i=1,\cdots, \dim \g_0\right> \oplus
\left<Y_a, \ a=1,\cdots, \dim \g_1\right>$ where $\g_0$ is a real
Lie algebra and $\g_1$ is a real representation of $\g_0$,
satisfying the following brackets

\beqa \label{flie} \left[X_i,X_j\right]&=& f_{ij}{}^k X_k, \ \
\left[X_i,Y_a\right]= R_{ia}{}^b Y_b, \\
\left\{Y_{a_1},\cdots,Y_{a_F}\right\}&=&
\sum \limits_{\sigma \in S_F}
 Y_{a_{\sigma(1)}} \cdots Y_{a_{\sigma(F)}}
= Q_{a_1 \cdots a_F}{}^iX_i, \nonumber
\eeqa

\noi
(where $S_F$ is the group of permutations of $F$ elements)
and fulfilling the following Jacobi identities for any $Y_{a_1},
\cdots, Y_{a_{F+1}}$ in $\g_1$

\beqa \label{Jac}
\sum \limits_{i=1}^{F+1} \left[ \left\{
Y_{a_1},\cdots,Y_{a_{i-1}},Y_{a_{i+1}},\cdots,Y_{a_{F+1}}
\right\},Y_{a_i}\right]
=0. \eeqa

\noi Looking at the various brackets, one immediately observes
that a Lie algebra of order $F$ is endowed with two different
products: one binary given by the usual commutators, and one of
order $F$ given by a fully symmetric product. Furthermore, a
direct inspection of \eqref{flie} and \eqref{Jac} shows that Lie
algebras of order $F$ are  $F-$ary extensions of Lie
superalgebras, where the anticommutator is replaced by a fully
symmetric  bracket of order $F$. Many examples of Lie algebras of
order $F$ were given in \cite{flie}, and a formal study of this
type of structure was initiated in \cite{gr, hopf, color}.

Subsequently, a program of investigation of higher orders
extensions of the Poincar\'e algebra, in the framework of Lie
algebras of order $F$, was undertaken. Among various
possibilities cubic and quartic extensions of the Poincar\'e algebra
have been defined \cite{flie}.
The cubic extension of the Poincar\'e algebra
$I\mathfrak{so}_3(1,3)= \g_0 \oplus \g_1$, with $\g_0= I\mathfrak{so}(1,3)=
\left<L_{\mu \nu} = -L_{\nu \mu}, \ P_\mu, 0 \le \mu, \nu \le 3
\right>$ generating the Poincar\'e algebra and $\g_1= \left<
V_\mu, 0 \le \mu \le 3 \right>$ being the vector representation,
is defined by the brackets

\begin{eqnarray}  \label{3poin}
\left[L_{\mu \nu }, L_{\rho \sigma}\right]&=& \eta_{\nu \sigma }
L_{\rho \mu }-\eta_{\mu \sigma} L_{\rho \nu} + \eta_{\nu
\rho}L_{\mu \sigma} -\eta_{\mu
\rho} L_{\nu \sigma},  \notag \\
\left[L_{\mu \nu }, P_\rho \right]&=& \eta_{\nu \rho } P_\mu
-\eta_{\mu \rho } P_\nu, \ \left[L_{\mu \nu }, V_\rho \right]=
\eta_{\nu \rho } V_\mu
-\eta_{\mu \rho } V_\nu, \ \left[P_{\mu}, V_\nu \right]= 0,   \\
\{ V_\mu, V_\nu, V_\rho \}&=& \eta_{\mu \nu } P_\rho + \eta_{\mu
\rho } P_\nu + \eta_{\rho \nu } P_\mu,  \notag
\end{eqnarray}

\noi where $\eta_{\mu \nu} = \text{diag}(1,-1,-1,-1)$ is the
Minkowski metric.

The quartic extensions of the Poincar\'e algebra are constructed
by considering two Majorana spinors. In the
$\mathfrak{sl}(2,\mathbb C) \cong \mathfrak{so}(1,3)$ notations of
dotted and undotted indices, a left handed spinor is given by
$\psi_L{}^\alpha$ and a a right-handed spinors by $\bar
\psi_R{}_{\dot \alpha}$.  The spinor conventions to raise/lower
indices are as follows $\psi_L{}_\alpha
=\varepsilon_{\alpha\beta}\psi_L{}^\beta$, $\psi_L{}^\alpha
=\varepsilon^{\alpha\beta}\psi_L{}_\beta$,
$\bar\psi_R{}_{\dot\alpha}=\varepsilon_{\dot\alpha
\dot\beta}\bar\psi_R{}^{\dot\beta}$, $\bar\psi_R{}^{\dot\alpha}
=\varepsilon^{\dot\alpha\dot\beta}\bar\psi_R{}_{\dot\beta}$ with
$(\psi_\alpha)^* =\bar\psi_{\dot\alpha}$, $\varepsilon_{12} =
\varepsilon_{\dot 1\dot 2}=-1$, $\varepsilon^{12} =
\varepsilon^{\dot 1\dot 2}=1$. The $4D$ Dirac matrices,  in the
Weyl representation,  are
\begin{eqnarray}
\label{eq:gamma}
\gamma_\mu =
 \begin{pmatrix}
 0&\sigma_\mu\\
 \bar\sigma_\mu&0
 \end{pmatrix},
\end{eqnarray}
with
$
 \sigma_{\mu\,\alpha\dot\alpha}=(1,\sigma_i ),
 \ \ \bar\sigma_\mu{}^{\dot\alpha\alpha}=
 (1,-\sigma_i),
$
%
where $\sigma_i, i=1,2,3$ are the Pauli matrices. With these
notations, we introduce two series of Majorana spinors
$Q^I{}_{\alpha}, \bar Q_I{}_{\dot \alpha}$  satisfying the
relation $(Q^I{}_{\alpha})^\dag= \bar Q_I{}_{\dot \alpha}$ and
define the quartic extensions of the Poincar\'e algebra (we only
give the quartic brackets) by

\beqa
\label{4poin}
 \left\{Q^{I_1}{}_{\alpha_1},Q^{I_2}{}_{\alpha_2},
Q^{I_3}{}_{\alpha_3},
Q^{I_4}{}_{\alpha_4}
\right\}&=&0 \nonumber \\
\left\{Q^{I_1}{}_{\alpha_1},Q^{I_2}{}_{\alpha_2},
Q^{I_3}{}_{\alpha_3},
\overline{Q}_{I_4}{}_{\dot \alpha_4}
\right\}&=& 2i\Big(\delta^{I_1}{}_{I_4} \varepsilon^{I_2 I_3}
\varepsilon_{\alpha_2 \alpha_3} \sigma^\mu_{\alpha_1 \dot \alpha_4}
+\delta^{I_2}{}_{I_4} \varepsilon^{I_1 I_3}
\varepsilon_{\alpha_1 \alpha_3} \sigma^\mu_{\alpha_2 \dot \alpha_4}  \\
&+&\delta^{I_3}{}_{I_4} \varepsilon^{I_1 I_2}
\varepsilon_{\alpha_1 \alpha_2} \sigma^\mu_{\alpha_3 \dot \alpha_4}\Big) P_\mu
\nonumber \\
\left\{Q^{I_1}{}_{\alpha_1},Q^{I_2}{}_{\alpha_2},
\overline{Q}_{I_3}{}_{\dot \alpha_3},
\overline{Q}_{I_4}{}_{\dot \alpha_4}
\right\}&=& 0,
\nonumber
\eeqa

\noi
the remaining
 brackets involving three $\bar Q$ and one $Q$ or four $\bar Q$ are obtained
by hermitian conjugation of the corresponding equation in
 \eqref{4poin} and $P_\mu^\dag = -P_\mu$\footnote{
With our conventions since there is no $i$ factors in the commutators,
we have that $P_\mu=\partial_\mu$, thus the
physical quadri-momentum is given by $-iP_\mu$.}. In
\eqref{4poin},
 $\varepsilon^{IJ}$ is the $SU(2)$ invariant tensor given by
$\varepsilon^{12}=-\varepsilon^{21}=1$.
This tensor enables us to define $SU(2)-$invariant, or equivalently
to raise or lower the indices. We define
$Q_I{}_\alpha= \e_{IJ} Q^J{}_\alpha$, with $\e_{IJ} \e^{JK}=\delta_I{}^K$.

The cubic extension \eqref{3poin}
 was intensively studied in the framework of
Quantum Field Theory \cite{cubic1, cubic2, noether,cours}.
However, it has been proven in \cite{cubic2} that  in four
dimensional space-time, and for a specific representation of
\eqref{3poin}, no-interacting terms are allowed. In order to get
more precise insight of the above higher order extensions, and to identify
interesting representations of \eqref{3poin} or \eqref{4poin}, one possible
direction would be to construct an adapted superspace, where the
algebra is realised upon differential operators.

\section{Superspace for higher order extensions of the Poincar\'e algebra}

In this section,  we construct an adapted superspace leading to an
{\it ad hoc} realisation of the algebras \eqref{3poin} and \eqref{4poin}.
More precisely, we introduce adapted variables such that the
transformations  generated by $V$ and $Q$ correspond to a translation
in some appropriate ``internal'' space hereafter   called
 superspace. We impose further that the algebra is
realised by means of  differential operators. The various
variables and differential operators, together with the basic
relations they have to satisfy, will be introduced  progressively.
It appears at the very end that the order two/three parafermions
turn out to be the relevant variables \cite{NE}. Moreover, it is
important to emphasise that the  parafermionic variables appear
quite naturally, but in a different way as they appeared
historically in the literature \cite{green,para}. Since the
construction is analogous for the cubic algebra \eqref{3poin} and
for the quadratic algebra \eqref{4poin}, in the next subsection we
construct the (ternary) superspace associated to \eqref{3poin}
with many details. Subsection \ref{quater} is devoted to the
analogous construction in the quartic case. However, it will be
shown that the latter case exhibits an exceptional behaviour, and
a quaternary superspace can even be constructed with  fermions.

\subsection{Ternary superspace}
\label{ternary}

A major progress towards the understanding of Lie algebras of
order three was achieved when it was realised that groups
associated to Lie algebras of order three may be defined
\cite{hopf}. Indeed, it was {\it a priori}  not obvious that
groups associated to ternary algebras can be defined, since for a
group the multiplication of two elements is always defined,
although for a ternary algebra only the multiplication of three
elements is defined. Moreover, the group structure enables us to
identify the parameters of the transformation \cite{hopf}. These
parameters turn out to be the natural cubic generalisation of the
Grassmann algebra (or the exterior algebra) called the
three-exterior algebra. This algebra is generated by four real
variables $\theta^\mu$ which are in the vector representation of
the Lorentz algebra and which satisfy the cubic relation \beqa
\label{roby} \left\{\theta^\mu,\theta^\nu,\theta^\rho\right\}=
\theta^\mu \theta^\nu \theta^\rho + \theta^\nu \theta^\rho
\theta^\mu + \theta^\rho \theta^\mu \theta^\nu + \theta^\mu
\theta^\rho \theta^\nu + \theta^\nu \theta^\mu \theta^\rho +
\theta^\rho \theta^\nu \theta^\mu =0. \eeqa

\noi This algebra can be defined over the real or complex fields.
For more details concerning the three-exterior algebra one
can see {\it e. g.} \cite{hopf}.
In the sequel we only consider the real three-exterior algebra.
This algebra was introduced long time ago by N. Roby \cite{roby},
and for that reason will call it  from now on the Roby algebra.
Since the transformation parameters belong to the Roby algebra, it
is natural to postulate that the superspace is generated by

\beqa \label{superespace} X=(x^\mu, \theta^\mu), \eeqa \noi where
$x^\mu$ belongs to the Minkowski space-time and $\theta^\mu$ are
the generators of the Roby algebra. This identification  is the
natural cubic extension of the superspace considered in usual
supersymmetric theories.

\bigskip

We further need to introduce the notion of ``conjugate'' momentum
of the variables $X$. To this extent, we have to identify some
derivative $\partial_\mu$ associated to the variables
$\theta^\mu$. To identify the action of $\partial_\mu$ on the
$\theta^\nu$, we assume further that a differential realisation of
the Lorentz algebra can be constructed from the variables $\theta$
and their derivative $
\partial$. We know that the variables $\theta$ are Lorentz
vectors. Following Green \cite{green,para}, the more general
quantisation which ensures that $\theta^\mu$ are vectors of the
Lorentz algebra is given by the parafermions \footnote{Or
parabosons, but the parabosonic algebra is incompatible with
requirement \eqref{roby}.}. We thus assume the parafermionic
relations\footnote{It should be noted that not all the relations
\eqref{para1} are independent, and some are related through
Jacobi identities.}

\beqa \label{para1}
\begin{array}{ll}
\left[\left[\theta^\mu,\theta^\nu\right],\theta^\rho\right]=0, &
\left[\left[\theta^\mu,\theta^\nu\right],\partial_\rho\right]=
-\delta^\mu{}_\rho \theta^\nu+\delta^\nu{}_\rho \theta^\mu, \\
\left[\left[\theta^\mu,\partial_\nu\right],\theta^\rho\right]=
\delta_\nu{}^\rho \theta^\mu, &
\left[\left[\theta^\mu,\partial_\nu\right],\partial_\rho\right]=
-\delta^\mu{}_\rho \partial_\nu,\\
\left[\left[\partial_\mu,\partial_\nu\right],\theta^\rho\right]=
-\delta_\mu{}^\rho \partial_\nu+\delta_\nu{}^\rho \partial_\mu
 &
\left[\left[\partial_\mu,\partial_\nu\right],\partial_\rho\right]=0.
\end{array}
\eeqa

\noi As a consequence, if we define \beqa \label{para4} {\cal
J}_{\mu \nu}= [\theta_\nu, \partial_\mu]-
[\theta_\mu,\partial_\nu], \eeqa

\noi the relations \eqref{para1} ensure that \eqref{para4} act
correctly on $\theta$ and $\partial$: \beqa [{\cal J}_{\mu \nu},
\theta_\rho]= \eta_{\nu
\rho} \theta_\mu - \eta_{\mu \rho} \theta_\nu. \eeqa

\noi Introducing further $P_\mu$, the conjugate momentum of
$x^\nu$ ($[P_\mu, x^\nu]= \delta_\mu{}^\nu$
), we thus
define the Lorentz generators as \beqa \label{lorentz} L_{\mu \nu}
= x_\nu P_\mu -x_\mu P_\nu +{\cal J}_{\mu \nu}. \eeqa

Since we are considering ternary algebras involving fully
symmetric products, putting \eqref{roby} and \eqref{para1}
together shows explicitly that we are considering parafermions of
order two. This means that the relations \eqref{roby} have to be
supplemented by \cite{para}\footnote{In the literature the
brackets \eqref{para1} and \eqref{para} are unified
$<\theta^\mu,\partial_\rho,\theta^\nu>=\delta^\mu{}_\rho
\theta^\nu+ \delta^\nu{}_\rho \theta^\mu, \
<\theta^\mu,\theta^\nu,\partial_\rho>= \delta^\nu{}_\rho
\theta^\mu$, where $<A,B,C>=ABC+CBA$ {\it etc}.} \beqa \label{para}
\left\{\theta^\mu,\theta^\nu,\theta^\rho\right\}&=&0, \nonumber \\
\left\{\theta^\mu,\theta^\nu,\partial_\rho\right\} &=&2
\delta^\mu{}_\rho \theta^\nu+
2 \delta^\nu{}_\rho \theta^\mu, \nonumber \\
\left\{\theta^\mu,\partial_\nu,\partial_\rho\right\}&=& 2
\delta^\mu{}_\nu \partial_\rho+
2 \delta^\mu{}_\rho \partial_\nu, \\
\left\{\partial_\mu,\partial_\nu,\partial_\rho\right\}&=&0.
\nonumber \eeqa

\noi It is interesting to observe that the construction leading to
\eqref{para} and \eqref{para1} goes in reverse order to that of
parafermionic algebras. Historically, parafermions were defined by
means of equation \eqref{para1}, in order to realise the Lorentz
algebra. After all the order of paraquantisation (here two, but
in general $p$) are specified by assuming on  which representation
of the Lorentz algebra the parafermionic algebra acts. Order $p$
parafermionic algebras involved fully symmetric brackets of order
$p+1$ and, in particular, order two parafermionic algebra give
rise to the brackets \eqref{para}. However, in our construction,
the cubic brackets \eqref{roby} are obtained from the very
beginning, by our superspace assumption. Finally, notice that the
order two parafermionic algebra  \eqref{para1}, \eqref{para} is a
non-faithful representation of the algebra \eqref{superespace}
since with respect to the Roby algebra
we have one more relation $[[\theta^\mu,\theta^\nu],\theta^\rho]=0$
\footnote{ In particular the Roby algebra is infinitely many generated
\cite{roby, hopf}
although the order three parafermionic algebra is finite dimensional
(see Section \ref{superfield}).
}.\\

However, we cannot  expect to construct a differential operator
$V_\mu$ from  $\partial_\mu$ and $\theta^\mu$ acting on
$\theta^\mu$ and satisfying the cubic relations \eqref{3poin}.
Indeed, the relations \eqref{para1} and \eqref{para} are cubic,
meaning that no bilinear relations upon $\theta^\mu$ and
$\partial_\nu$ are given and consequently no action of
$\partial_\mu$ on $\theta^\nu$ is specified\footnote{ It is
possible, however, to obtain matrix representations of
\eqref{3poin} by postulating cubic relations analogous to
\eqref{para} of \cite{cubic1}.}. This situation is very similar to
the implementation of the Noether theorem within the framework of
ternary symmetries, where the conserved charges generate the
symmetry through quadratic relations using the usual quantisation
procedure ({\it e.g.} the equal-time (anti)-commutation relations)
\cite{cubic1,noether, cours}.  We have shown in
\cite{cubic1,noether} that if we have an invariant Lagrangian
${\cal L}(\Phi)$ (where $\Phi$ is  a given multiplet of the algebra
\eqref{3poin}) with
conserved charges $\hat L_{\mu \nu}, \hat P_\mu, \hat V_\mu$ such
that after quantisation we have 
 the transformation laws $ [\hat L_{\mu \nu}, \Phi], [\hat P_{\mu},
\Phi], [\hat V_{\mu }, \Phi], $ the algebra is realised through
multiple-commutators

\beqa \label{adj} [\hat V_\mu,[\hat V_\nu, [\hat V_\rho,\Phi]]] +
\text{perm.}=
 \eta_{\mu \nu} [\hat P_\rho, \Phi]+
\eta_{\nu \rho} [\hat P_\mu,\Phi] + \eta_{\mu \rho}  [\hat
P_\nu,\Phi]. \eeqa

\noi This procedure is standard in the implementation of Lie
(super)algebra in Quantum Field Theory, but the equation
corresponding to \eqref{adj} in this case is not the end of the
story  since the Jacobi
identities allow to obtain a relation which is independent of the
fields $\Phi$. But here, in the context of ternary symmetries, the
situation is very different, since the Jacobi identities
\eqref{Jac} do not allow to write the algebra in a $\Phi$
independent form. This weaker realisation of the algebra has the
interesting consequence that it enables us to  consider algebraic
structure (in Quantum Field Theories), different from Lie
superalgebras,  without contradicting  the spin-statistics theorem
(see \cite{cours} for a discussion). Finally, it is a matter of
calculation to check that the Jacobi identities are satisfied by
the realisation \eqref{adj}.
\\

As we have recalled briefly, the implementation of the Noether
theorem in higher order algebras automatically leads to an
algebraic realisation through multiple commutators. Since for the
parafermionic algebra, the natural action is defined also by means
of commutators (see \eqref{para1}), it is tempting to try to
define a superspace in which the algebra is realised in the form
of \eqref{adj}. Then, the  parafermionic algebra will be the
cornerstone of the realisation of the algebra \eqref{3poin} on the
superspace $X=(x^\mu, \theta^\mu)$. We then introduce the
parameters of the transformations $\e^\mu$ such that we have the
transformation
$$
\theta^\mu \to \theta'{}^\mu=\theta^\mu + \e^\mu,
$$
\noi under \eqref{3poin}. This means that the variables $\theta'$
are of the same type as the variables $\theta$. So do the
variables $\e$.  Now we would like to define the generators of our
algebra. The algebra \eqref{3poin} is cubic and the  variables
$\e$ satisfy also the cubic relations \eqref{para1} and
\eqref{para}. It is known that in general the tensor product of
two algebras has no meaning (and this is even a more difficult
task for  algebras defined by cubic relations like \eqref{3poin}
and \eqref{para}, \eqref{para1}). We thus assume that the
variables $\e$ and the generators of the transformation are {\it
indissociable}. Since for the para-Grassmann algebra the natural
objects are the commutators, it is natural to define $V$ by means
of a commutator. There are two parts in the generators $V$. One
leading to the transformations on the variables $\theta$ and one
transforming the variables $x$: $V=V_\theta + V_x$ such that
$[V_\theta,x]=0$ and $[V_x,\theta]=0$. Having only the variables
$\theta^\mu$ and the parameters $\e^\mu$ it is not difficult to
observe that it is not possible to define a $V_x$ commuting with
the $\theta$'s and being a Lorentz scalar. We thus introduce one
more parameter $\theta$  which is a paragrassmann variable in the
scalar representation of the Lorentz algebra\footnote{This
additional variable can be understood as coming from a
compactification of the $(1+4)D$ to the $(1+3)D$  Minkowski
space-time. Thus $(\theta^\mu,\theta)$ is in the vector
representation of $SO(1,4)$.} such that
 \beqa \label{rep}
V=\left[\e^\mu,\partial_\mu\right]+ \left[\theta,\theta^\mu\right]
\left[\e^\sigma, \theta_\mu\right]P_\sigma, \eeqa

\noi
which gives \beqa \label{trans} \delta
\theta^\alpha=\left[V,\theta^\alpha\right]=\e^\alpha, \ \ \delta
x^\alpha=\left[V,x^\alpha\right]= \left[\theta,\theta^\mu\right]
\left[\e^\alpha, \theta_\mu\right]. \eeqa \noi It is important to
realize that,  the $\delta x^\alpha$ are commuting real variables.



Having constructed a differential realisation of the cubic
extension of the Poincar\'e algebra \eqref{3poin}, we are now
looking for the closure of the algebra. We have to check the
algebra in the sense of \eqref{adj} on any monomial in $\theta$.
In particular, we have

\beqa \label{leib}
[V_1,[V_2,[V_3,\theta^{\alpha_1}\theta^{\alpha_2}
\theta^{\alpha_3}]]]&=& \e_1^{\alpha_1}  \e_2^{\alpha_2}
\e_3^{\alpha_3} + \e_2^{\alpha_1}  \e_3^{\alpha_2} \e_1^{\alpha_3}
+ \e_3^{\alpha_1}  \e_1^{\alpha_2}   \e_2^{\alpha_3}\nonumber \\ &
+& \e_1^{\alpha_1}  \e_3^{\alpha_2}   \e_2^{\alpha_3} +
\e_2^{\alpha_1}  \e_1^{\alpha_2}   \e_3^{\alpha_3} +
\e_3^{\alpha_1}  \e_2^{\alpha_2}   \e_1^{\alpha_3}, \eeqa

\noi which is fully symmetric in the indices $1,2,3$. This means
that $[V_1,[ V_2,[ V_3,\theta^{\alpha_1}\theta^{\alpha_2}
\theta^{\alpha_3}] + \text{perm}$,  never vanishes (in order to
simplify the notations, we denote $V_1.V_2.V_3.
\theta=[V_1,[V_2,[V_2,\theta]]]$ and $\{V_1,V_2,V_3\}. \theta =
[V_1,[V_2,[V_2,\theta]]] + \text{perm.}$ {\it etc.}). { It turns
out that this problem is independent of the realisation of the
operator $V$. Actually, if we simply assume to have an operator
$\delta$ such that $\delta . \theta = \e$, together with the
Leibniz rule, we automatically have for $\delta_1 \delta_2
\delta_3 (\theta^{\alpha_1}\theta^{\alpha_2} \theta^{\alpha_3})$
the R.H.S. of \eqref{leib}. In fact this discrepancy is already
present in supersymmetric theories. But in this case, since the
parameters are anticommuting Grassmann parameters, the natural
solution is to replace anticommutators by commutators when the
parameters are taken into account. In our case we have some
analogy to the procedure above. Recall that the formal study of
Lie algebras of order three \cite{hopf} leads naturally to a
$\mathbb Z_3-$twisted tensor product. This means, in particular,
that if we consider three successive transformations $\e_1, \e_2,
\e_3$, we have a $\mathbb Z_3 \times \mathbb Z_3 \times \mathbb
Z_3$ graded structure. As a consequence, the algebraic structure
which emerges from this grading is a ternary analog of the colour
algebras were the bracket is no more fully symmetric
\cite{coloralg}\footnote{ Colour algebras were introduced as a
possible generalisation of Lie (super)algebras, were the brackets
are neither symmetric nor antisymmetric. In the various brackets
the usual plus or minus sign of the (anti)commutators is
substituted by a commutation factor. }. This means that, as in the
case of usual Lie algebras, the introduction of the parameters of
the transformation forces us to consider a different, but related
algebraic structure. This can be seen as some analogy to the
Jordan-Wigner transformations in that context. This is the inverse
process of the decoloration theorem proved in \cite{color}, which
states that a ternary colour algebra is isomorphic to a Lie
algebra of order three. Ternary colour algebras have been studied
in \cite{color}. The basic tool to define colour algebras is a
grading determined by an Abelian group. Here the grading group is
given by $\mathbb Z_3 \times \mathbb Z_3 \times \mathbb Z_3$. The
latter, besides defining the underlying grading in the structure,
moreover provides a new object known as commutation factor defined
by

\beqa
\label{3-com}
N(\vec a, \vec b)= q^{a_1(b_2 + b_3) + a_2 b_3 -
 b_1(a_2 + a_3) -b_2 a_3},
 \eeqa

\noi
where $q=e^{\frac{2i \pi}{3}}$ and $\vec a, \vec b \in \mathbb Z_3^3$.
The trilinear
bracket  is now defined by\footnote{There is also some additional Jacobi
identities which are however not relevant for our analysis.}

\beqa
\label{3-color}
\lb V_1, V_2, V_3 \rb &=& V_1 V_2 V_3 +
N\Big(\gr(\e_1),\gr(\e_2)+\gr(\e_3)\Big) V_2 V_3 V_1
 \\
&+&
 N\Big(\gr(\e_1)+\gr(\e_2),\gr(\e_3)\Big)  V_3 V_1 V_2   \nonumber \\
&+& N\Big(\gr(\e_2),\gr(\e_3)\Big) V_1 V_3 V_2+
 N\Big(\gr(\e_1),\gr(\e_2)\Big)V_2 V_1 V_3 \nonumber \\
&+& N\Big(\gr(\e_1),\gr(\e_2)\Big) N\Big(\gr(\e_1),\gr(\e_3)\Big)
N\Big(\gr(\e_2),\gr(\e_3)\Big)V_3 V_2 V_1. \nonumber \eeqa

\noi In our case, with the commutation factor given by
\eqref{3-com}, and  $\text{gr}(\e_1)=(1,0,0),\;
\text{gr}(\e_2)=(0,1,0),\; \text{gr}(\e_3)=(0,0,1)$, the cubic
brackets adopt the following form

\beqa
\label{col3}
\lb V_1, V_2, V_3 \rb
&=&V_1 V_2 V_3 +  q^2 V_2 V_3 V_1 + q^2 V_3 V_1 V_2 + q V_1 V_3
V_2 + q V_2 V_1 V_3 + V_3 V_2 V_1. \eeqa

\noi In particular, since the constraint $1+q+q^2=0$ is satisfied
and $V_1 .V_2 .V_3.(\theta^{\alpha_1} \cdots \theta^{\alpha_n})$
is fully symmetric in the subindices $1,2,3$, we automatically
have that
$$\lb V_1, V_2, V_3 \rb.(\theta^{\alpha_1} \cdots \theta^{\alpha_n})
=0.$$

\noi Performing a similar computation for the space-time
coordinates, we obtain the identities \beqa \label{deltax} \lb
V_1,V_2,V_3 \rb . x^\alpha &=&
-q^2[\theta,\e_2^\mu][\e_3^\alpha,\e_1{}_\mu]
-q^2[\theta,\e_1^\mu][\e_3^\alpha,\e_2{}_\mu ]\nonumber \\
&&-[\theta,\e_2^\mu][\e_1^\alpha,\e_3{}_\mu]
-[\theta,\e_3^\mu][\e_1^\alpha,\e_2{}_\mu]\nonumber \\
&&-q[\theta,\e_1^\mu][\e_2^\alpha,\e_3{}_\mu]
-q[\theta,\e_3^\mu][\e_2^\alpha,\e_1{}_\mu] = a^\alpha . \eeqa

\noi It is important to notice that the $a^\beta$  are complex
numbers. This means that the ``coloration'' of the algebra
$I\mathfrak{so}_3(1.3)$, coming from our adapted Jordan-Wigner
transformation gives rise to the algebra \eqref{col3}, which is
manifestly a complex algebra since the structure constants are
complex \footnote{
In fact since  $\mathbb Z_3^3$ is complex, the grading makes sense only in
a complexification $I\mathfrak{so}_3(1.3) \otimes_{\mathbb R} \mathbb C$
of the cubic algebra $I\mathfrak{so}_3(1.3)$.
}. This deserves some explanation. The $\e$ are real
parafermions, therefore the transformation properties
\eqref{trans} ensure that $\delta x$ and $\delta \theta$ are both
real. However, since $a^\beta$ is complex, this means that the
cubic algebra \eqref{3poin} is {\it realised in a
complexification} of the superspace $(x,\theta)$. In other words,
the algebra cannot be realised on a real vector space. This is the
best possible result in this direction using this ansatz.

\subsection{Quaternary superspace}
\label{quater} The construction of quaternary superspaces goes
along the same lines of the construction of cubic superspace, but
with  some differences that we give now. Firstly the internal
variables are given by order three parafermions instead of order
two parafermions.  Furthermore these variables are in the spinor
representations of the Lorentz algebra
(we have two Majorana spinors): $\theta^{I}{}_{\alpha},
\bar \theta_I{}_{\dot \alpha}$ with $\theta^{I}{}_{\alpha}{}^\dag
= \bar  \theta_I{}_{\dot \alpha}$. Denoting generically by
$\theta^a$ and $\partial_a$ the order three parafermions and their
associated momenta, the order three parafermionic algebra is then
given by \eqref{para1} and  the quartic relations

\beqa
\label{4para}
\left\{\theta^{a_1},\theta^{a_2},\theta^{a_3},\theta^{a_4}
\right\}&=&0, \nonumber \\
\left\{\theta^{a_1},\theta^{a_2},\theta^{a_3},\partial_{a_4}
\right\}&=&
5\delta^{a_1}{}_{a_4}\left\{\theta^{a_2},\theta^{a_3}\right\} +
5\delta^{a_2}{}_{a_4}\left\{\theta^{a_1},\theta^{a_3}\right\} +
5\delta^{a_3}{}_{a_4}\left\{\theta^{a_1},\theta^{a_2}\right\},
\\
\left\{\theta^{a_1},\theta^{a_2},\partial_{a_3},\partial_{a_4}
\right\}&=& 5 \delta^{a_1}{}_{a_3}
\left\{\theta^{a_2},\partial_{a_4}\right\} + 5
\delta^{a_1}{}_{a_4} \left\{\theta^{a_2},\partial_{a_3}\right\}+5
\delta^{a_2}{}_{a_3} \left\{\theta^{a_1},\partial_{a_4}\right\}
\nonumber \\
&+& 5 \delta^{a_2}{}_{a_4}
\left\{\theta^{a_1},\partial_{a_3}\right\} -\frac92
\delta^{a_1}{}_{a_3} \delta^{a_2}{}_{a_4}- \frac92
\delta^{a_1}{}_{a_4} \delta^{a_2}{}_{a_3}, \nonumber \eeqa

\noi
plus similar relations involving one $\theta$ and three $\partial$ or
four $\partial$.

Introducing the parameters of the transformation, the quartic supercharges
are now given by

\beqa
\label{4charge}
Q=&&[\e^I{}^\alpha, \partial_I{}_{\alpha}] + 2i
[\e^I{}^{\alpha},\theta_I{}_{\alpha}]
[\theta^J{}^\beta,\bar \theta_J{}^{\dot \beta}] \sigma^\mu{}_{\beta \dot \beta}
P_\mu, \\
\bar Q=&-&[\bar \partial^I{}_{\dot \alpha},\bar \e_I{}^{\dot \alpha}] - 2i
 [\theta^J{}^\beta,\bar \theta_J{}^{\dot \beta}]
[\bar \theta^I{}_{\dot \alpha},\bar \e_I{}^{\dot \alpha}]
 \sigma^\mu{}_{\beta \dot \beta}
P_\mu.\nonumber \eeqa

\noi leading to the transformations 

\beqa
\label{4trans}
&\delta x^\mu= [Q,x^\mu]+ [\bar Q,x^\mu]=
2i\Big([\e^I{}^{\alpha},\theta_I{}_{\alpha}]
[\theta^J{}^\beta,\bar \theta_J{}^{\dot \beta}]
-[\theta^J{}^\beta,\bar \theta_J{}^{\dot \beta}]
[\bar \theta^I{}_{\dot \alpha},\bar \e_I{}^{\dot \alpha}]
 \Big) \sigma^\mu{}_{\beta \dot \beta}
 , \nonumber \\
&\delta \theta^I{}_\alpha=[Q, \theta^I{}_\alpha]=
 \e^I{}_\alpha, \ \ \
\delta \bar \theta_I{}_{\dot \alpha}=
[\bar Q,\bar \theta_I{}_{\dot \alpha}]=
\bar \e_I{}_{\dot \alpha}.
\eeqa

\noi
A direct inspection shows that $\delta x^\mu$ is real.

Finally, as in the cubic case, the algebra is realised in a
complexification of the quaternary superspace, but now with a
$\ZZ_4 \times \ZZ_4 \times \ZZ_4 \times \ZZ_4-$grading, where the
corresponding commutation factor is given by

\beqa
\label{4-mut}
N(\vec a, \vec b)=i^{a_1(b_2+b_3+b_4)+a_2(b_3+b_4)+a_3 b_4-
b_1(a_2+a_3+a_4)-b_2(a_3+a_4)-b_3 a_4}.
\eeqa

We consider then four successive
transformations $Q_1,\cdots,Q_4$ with grading
 $\gr(\e_1)=\gr(\bar \e_1)=(1,0,0,0),\
\gr(\e_2)=\gr(\bar \e_2)=(0,1,0,0),\
\gr(\e_3)=\gr(\bar \e_3)=(0,0,1,0),\
\gr(\e_4)=\gr(\bar \e_4)=(0,0,0,1)$ and define

\beqa
&&\lb Q_1,Q_2,Q_3,Q_4\rb=Q_1 \lb Q_2,Q_3,Q_4 \rb +
N(\gr(\e_1),\gr(\e_2)) Q_2 \lb  Q_1,Q_3,Q_4  \rb  \nonumber\\
&&
N(\gr(\e_1)+\gr(\e_2),\gr(\e_3)) Q_3 \lb  Q_1,Q_2,Q_4 \rb +
N(\gr(\e_1)+\gr(\e_2)+\gr(\e_3),\gr(\e_4)) Q_4 \lb  Q_1,Q_2,Q_3 \rb
\nonumber \\
&&=Q_1\lb Q_2,Q_3,Q_4\rb + i Q_2\lb  Q_1,Q_3,Q_4  \rb -Q_3 \lb
Q_1,Q_2,Q_4 \rb -i Q_4 \lb  Q_1,Q_2,Q_3 \rb,  \eeqa

\noi with $\lb Q_i, Q_j,Q_k\rb$ given by \eqref{3-color}, but with
commutator factor \eqref{4-mut} instead of \eqref{3-com}. As in
section \ref{ternary}, we can easily check that $\lb
Q_1,Q_2,Q_3,Q_4\rb$ vanishes on any polynomial on $\theta$ and
generate a space-time translation with a complex parameter.

\subsection{Fermions for higher order superspaces}
In principle, the variables $\theta$ should satisfy the Roby
algebra $\{\theta,\theta,\theta\}=0$ for cubic extensions and the
Roby algebra $\{\theta,\theta,\theta,\theta\}=0$ for quartic
extensions, with no more relations among the $\theta$'s
but $[[\theta,\theta],\theta]=0$. However,
if we ``relax'' these assumptions, this gives rise to the
possibility to realise higher order algebras using only fermions.
This realisation can be put on the same footing with the
realisation of the algebra in relation with the Noether theorem,
where the algebra is realised by means of bosons and fermions. The
only relevant relations in the construction of ternary  (resp.
quaternary) superspaces are equations \eqref{para1} and
$\{\theta,\theta,\theta\}=0$ (resp.
$\{\theta,\theta,\theta,\theta\}=0$). It turns out that usual
fermions do satisfy \eqref{para1} and \eqref{para} or
\eqref{4para} (but with a different normalisation for the last
equations). This leads to an alternative construction of higher
order superspaces, with usual fermions instead of parafermions. Of
course that in the cubic case, this possibility is excluded since
the generators $V_\mu$ are in the vector representation of the
Lorentz algebra. Indeed, in this case superfields (see Section
\ref{superfield}) will automatically generate commuting fermions
or anticommuting bosons. However, for extensions involving
spinors this possibility is still open. If one considers the
quartic algebra \eqref{4poin}, it is straightforward to verify
that it admits generalisation in any space-time dimensions. As
happens for supersymmetric theories, these extensions depend on
the space-time dimension and of the properties of spinors.  As we
now show, there is one space-time dimension where these quartic
algebras present some exceptional features, namely $D=10$.

Before giving the analog of \eqref{4poin} in  ten space-time
dimensions, lets us recall some properties of spinors which are
useful for us. (For more details one can see {\it e.g.}
\cite{cliff}.) Dirac spinors can be defined in any space-time
dimensions and Weyl spinors in even space-time dimensions. A
Majorana (or pseudo-Majorana) spinor,
 that is a real spinor, can be defined in $D=4,8,9,11$ (mod. $8$)
and a Majorana
Weyl spinor is only defined in  $D=10$ (mod. $8$). In this brief
section, we are mainly interested in the minimal quartic
extensions, that is, when real spinors (or real Weyl spinors)
do exist, thus when
$D=9,10$ or $D=11$. Furthermore, since our extensions involve
fully symmetric product we consider the case where the charge
conjugation matrix is symmetric. If $\Gamma^M$ are the
$\gamma-$matrices in $D-$dimensions and $C$ the charge-conjugation
matrix defined by

\beqa
\label{C}
(\Gamma^M)^t= \pm C \Gamma^M C^{-1},
\eeqa

\noi
where $M^t$ denotes the transpose of the matrix $M$.
Both signs are possible when $D$ is even and the sign
of the R.H.S. is fixed when $D$ is odd. For $D=9$, the matrices
$C$ and $\Gamma^M C$ are both symmetric ($C$ is denoted $C_+$)
although in $D=10$, there is one choice of $C$ (denoted $C_+$),
that ensures that $C_+$ and $\Gamma^M C_+$ are both symmetric. The
other choice of $C$, (denoted $C_-$), implies that we have $C_-$
antisymmetric and $\Gamma^M C_-$ symmetric. Thus when $D=9,10$
Majorana (Weyl) spinors exist together with a symmetric charge
conjugation matrix.

Now we are ready to give the quartic extension of the Poincar\'e
algebra. As mentioned previously these extensions exist in any
space-time dimensions, but due to the special properties of
Dirac-matrices in $D=9,10$, these extensions are simpler in these
cases. Since they present some analogies, they will be presented
on the same footing. Consider $Q_A$ a Majorana spinor   and
$\Gamma^M,$ $M=0,\cdots,8$ (resp. $M=0,\cdots,9$) the Dirac
$\Gamma-$matrices in $D=9$ (resp. $D=10$)\footnote{ In $D=10$ the
charge conjugation matrices $C_+$ or $C_-$ connect left-handed to
right-handed spinors. This means that we cannot consider a quartic
extension of the Poincar\'e algebra with only a Weyl spinor. This
can be seen in an equivalent way. Indeed, if $\psi_L$ denotes a
left-handed spinor, we have that $\psi_L \otimes \psi_L= [1]\oplus
[3] \oplus [5]_+$ with $[p]$ a $p-$form and $[5]_+$ a self-dual
five-form.}.
 The quartic extension analogous to \eqref{4poin}
is given by

\beqa
\label{D=9-10}
\{Q_{A_1}, Q_{A_2},Q_{A_3},Q_{A_4}\}&=&
C_+{}_{A_1 A_2} (\Gamma^MC_+)_{A_3 A_4} P_M +
C_+{}_{A_1 A_3} (\Gamma^MC_+)_{A_2 A_4} P_M  \nonumber \\
&+& C_+{}_{A_1 A_4} (\Gamma^MC_+)_{A_2 A_3} P_M  +
C_+{}_{A_2 A_3} (\Gamma^MC_+)_{A_1 A_4} P_M \\
&+& C_+{}_{A_2 A_4} (\Gamma^MC_+)_{A_1 A_3} P_M + C_+{}_{A_3 A_4}
(\Gamma^MC_+)_{A_1 A_2} P_M. \nonumber \eeqa

\noi Among these two extensions, the $D=10$ quartic extension of
the Poincar\'e algebra is very special. There exist two charge
conjugation matrices $C_+$ and $C_-$ with different properties of
symmetry. As in the preceding subsection, one can define a
quaternary supercharge, considering order three parafermions in
the Majorana representation of $SO(1,9)$. If we denote by
$\theta^A,\partial_A$ the order three parafermions and by
$\epsilon^A$ the parameters of the transformation,
 we have that

\beqa
\label{paraD=10}
Q=[\epsilon^A,\partial_A]+
C_-{}_{BC}[\theta^B, \theta^C] \  (\Gamma C_+)_{AD}[\epsilon^A, \theta^D].
\eeqa

However, in this case, one is also able to construct a quaternary
supercharge with usual fermions Introducing $\psi^A, \partial_A$
real Grassmann variables (in the Majorana representation of
$SO(1,9)$) and their associated derivative we can construct in
$D=10$ the quaternary supercharges using only fermions

\beqa
\label{superD=10}
Q_A= \partial_A + (C_-{}_{BC}\psi^B \psi^C) \  (\Gamma C_+)_{AD} \psi^D.
\eeqa

\noi Of course, as happened in Section \ref{quater}, the algebra
is realised with multiple commutators in a complexification of the
superspace. It is important to point out that this realisation is possible
due to the special properties of the $C_\pm-$matrices in ten space-time
dimensions.
Finally, looking to the algebraic structure
\eqref{D=9-10} and its differential realisation \eqref{superD=10}
one may wonder whether this quartic relation is related with type
IIA supersymmetry \cite{IIA}?

\bigskip
To conclude this section let us mention that similar (complex)
variables were used in \cite{kern} in a different context. In
addition, order two paragrassmann algebras were used for the
description of a superspace associated to parasupersymmetric
extensions of the Poincar\'e algebra in \cite{parag}.

\section{Superfields}
\label{superfield}

In the previous section we were able to construct an adapted
superspace, where the underlying algebra \eqref{3poin} and
\eqref{4poin} are realised in a differential way. Considering
functions of the variables \eqref{superespace} (and its quartic
analogue)
 gives us the opportunity to obtain various
representations of the algebra \eqref{3poin} or \eqref{4poin}.
In analogy with
supersymmetric theories, these functions will be called
superfields. Since the definition of superfields is analogous in
the ternary and quaternary cases, we study the first case with
many details, the second being obtained in straightforward manner.
In addition, we will analyze the extremely important problem of
the existence of covariant derivatives. Although some serious
obstruction exist in the full general case, slight modifications
will enable us to find operators $D$ acting like covariant
derivatives, in both the ternary and quartic cases. 
However, it turns out that
we obtain, with this modification,
 an interesting effect, namely, space-time translations
of null vectors.
\\

\subsection{Ternary superfields}
In what follows, we define a scalar superfield, {\it i.e.}, a
superfield invariant under Lorentz transformations. We then
identify the various representations of the Lorentz algebra
together with its spin content, which appears in the decomposition
of the scalar superfield. We also analyse the possibility to
construct another type of superfields using the so-called Green
ansatz.\newline A scalar superfield is given by a function \beqa
\label{superPhi} \Phi(x,\theta). \eeqa

\noi Developing \eqref{superPhi} with respect to the $\theta$, we
get monomials in the parafermionic variables $\theta$. The first
of the relations in \eqref{para1}  and \eqref{para}, respectively,
imply the identity

\beqa \label{cub} \theta^\mu \theta^\nu \theta^\rho + \theta^\rho
\theta^\nu \theta^\mu=0. \eeqa

\noi Upon successive application of the identity \eqref{cub}, we
get the relation $(\theta^\mu \theta^\nu
\theta^\rho)^3=0.$ It can be shown that the relations
\eqref{cub} ensure that we have a finite number of monomials in
the development of the superfield $\Phi$. To develop equation
\eqref{superPhi}, we have to identify its components with respect
to the Lorentz group. A series of results upon parafermions, that
we recall now,
 has been established. Given an arbitrary
monomial of degree $n$, say $\theta^{\alpha_1} \cdots
\theta^{\alpha_n}$, it can be written as a linear combination of
terms of the shape (see \cite{para, theo1} for details on the
method)

\beqa \label{theo1}
&[\theta^{\mu_1},\theta^{\nu_1}]\cdots[\theta^{\mu_{p}},\theta^{\nu_p}]
\{\theta^{\rho_1},\cdots,\theta^{\rho_q}\}, \\
&2p+q=n, 0\le q \le 2, \ 0 \le \alpha^{\mu_1},\cdots,
\alpha^{\mu_p},\alpha^{\nu_1},\cdots, \alpha^{\nu_p},
\alpha^{\rho_1},\cdots,\alpha^{\rho_q} \le 3. \nonumber \eeqa

The second result is even more interesting for our approach. It is
known that an arbitrary tensor of order $n$ decomposes into
irreducible representations of the Lorentz group characterised by
a certain Young tableau \cite{young}. Moreover, multiplicity of
representations in the above decomposition is generally greater
than one. However, for order two parafermions \eqref{para1},
\eqref{para} the situation changes drastically. If we define $
{\cal P}_n = \Big\{ \theta^{\alpha_1} \cdots \theta^{\alpha_n}, \
\ 0 \le \alpha_1,\cdots, \alpha_n \le 3 \Big\}, $ it turns out
that ${\cal P}_n$ contains {\it one and only one} irreducible
subspace corresponding to each possible Young diagram of $n$
squares whose first row consists of not more than $2$ squares
\cite{para,theo2}\footnote{\label{paraf}Analogous properties are equally valid
for order $p$ parafermions.}. This  property of multiplicity of
equivalent representations is one of the most appealing
characteristics of parafermions. Technically, since to each
allowed Young diagram there corresponds only one irreducible
representation, we are able to chose the Young tableaux which
gives us the more convenient results.

\smallskip

We will not give a systematic study of the decomposition of the
superfield $\Phi$ upon the various Young tableaux, since these
computations are straightforward but lengthy. We only give the
general method to identify the spin content of the fields
appearing in the decomposition of the superfield $\Phi$. We first
recall that, in order to identify a representation associated to a
given Young tableau, one as to define the Young symmetriser
associated to the considered Young tableau. It is written as
$P=SA$, {\it i.e.}, we first antisymmetrise the columns and then
symmetrise the rows. As an illustration,
 we give one example. Consider the representation
associated to the Young tableau\footnote{If along the same lines, one
calculates the representation associated to the Young tableau
{\tiny{$
\begin{tabular}{|c|c|}
\hline
1&3 \\
\hline
2\\
\cline{1-1}
\end{tabular}$}} one obtains $\frac13[\theta^{M_3},
[\theta^{M_1},\theta^{M_2}]]$, which vanishes. }

$$
\small{
\begin{tabular}{|c|c|}
\hline
1&2 \\
\hline
3\\
\cline{1-1}
\end{tabular}
}
$$

\noi
its Young symmetriser is given by $P=\frac13\big(1+(12)\big)\big(1-(13)\big)$
(where $(ab)$ means the transposition of $a$ and $b$) and correspondingly it
leads to the representation\footnote{In what
follows, the letters $M,N =0,\cdots,4$
correspond  to the $5-$dimensional indices and
we have identified
$\theta= \theta^4$.}

$$
P \theta^{M_1} \theta^{M_2} \theta^{M_3}=\frac13\{\theta^{M_1},\theta^{M_2}\}
\theta^{M_3} -\frac13 \{\theta^{M_2},\theta^{M_3}\}
\theta^{M_1}.
$$

\noi Proceeding along the same lines, we obtain 17
representations, varying from the trivial representation to the
one-dimensional representation specified to the Young tableau

{\small{
$$
\begin{tabular}{|c|c|}
\hline
&\\
\hline
&\\
\hline
&\\
\hline
&\\
\hline
&\\
\hline
\end{tabular}\ .
$$
}}

\noi The tensors appearing in this decomposition can be seen to
constitute irreducible representations of $GL(1,4)$, thus to
identify representations of $SO(1,3)$, we first have to identify
its $GL(1,3)$ content and then we have to extract traceless
tensors. For instance, the $GL(1,4)$-tensor
$\{\theta^M,\theta^N\}$ gives rise to the $GL(1,3)$ tensors
$\{\theta^\mu,\theta^\nu\}$
 and $\{\theta^\mu,\theta\}$.
The first one leads then to
 $\theta^\mu
\theta_\mu$ and $\{\theta^\alpha,\theta^\beta\}-\frac12 \theta^\mu
\theta_\mu \eta^{\alpha \beta}$.
Having identified the $SO(1,3)$ content of the decomposition of the
various fields,
we can now identify the spin or
the helicity content of the various representations.
 For the former identification, corresponding to massive
particles, we have to study the embedding $\mathfrak{so}(3)
\subset \mathfrak{so}(1,3) $, although for massless particles, we
have to decompose the various tensors according to the reduction
chain $\mathfrak{so}(2) \subset \mathfrak{so}(1,3)$. Considering
only non-isomorphic representations of $\mathfrak{so}(1,3)$,
we obtain for the massive
case

\beqa \label{spin}
\begin{array}{cclcclccl}
{
\begin{tabular}{|c|}
\hline
\\
\hline
\end{tabular}}&=&\ \utilde{\mathbf{1}} \oplus \utilde{\mathbf{3}}, &
{
\begin{tabular}{|c|c|}
\hline
&\\
\hline
\end{tabular}}&=&\utilde{\mathbf{1}} \oplus \utilde{\mathbf{3} } \oplus
(\utilde{\mathbf{5}}
\oplus \utilde{\mathbf{1}}), &{
\begin{tabular}{|c|c|}
\hline
& \\
\hline
\\
\cline{1-1}
\end{tabular}}
&=&2 \times  \utilde{{\bf 5}} \oplus 3 \times \utilde{{\bf 3}} \oplus
\utilde{{\bf 1}},
  \\
 \\
{
\begin{tabular}{|c|c|}
\hline
&\\
\hline
&\\
\hline
\end{tabular}}&=&3\times\utilde{{\bf 5}} \oplus \utilde{{\bf 3}} \oplus 2
\times\utilde{ {\bf 1}},
&{
\begin{tabular}{|c|}
\hline
\\ \hline \\
\hline
\end{tabular}}&=&2 \times \utilde{\mathbf{3}},
\\
\end{array}
\eeqa

\noi where the representation of dimension $2s+1$ corresponds to a
particle of spin $s$.

Thus a general superfield decomposes into the representations
obtained in the process above, and corresponds to a given
representation of the complexification of the algebra
\eqref{3poin}. It turns out that this representation is reducible.
Furthermore,  as it is the case in usual supersymmetry, it should
certainly be interesting to construct constraint superfields in
order to obtain various models invariant under equations
\eqref{3poin}. Having the decomposition of the field $\Phi$ into
the monomials in $\theta$, and using \eqref{trans}, we are in
principle able to obtain the transformation properties for the
various components of $\Phi$.
 \\

At this point, an interesting additional possibility to construct
slightly different superfields emerges naturally. It is well known
that a physically important representation of the parafermionic
algebra exists, namely the Green ansatz given by

\beqa
\label{green} \theta^M = \frac{1}{\sqrt{2}}
\left(\theta^M_{(1)} + \theta^M_{(2)}\right), \ \
\partial_M = \frac{1}{\sqrt{2}}
\left(\partial_M^{(1)}+\partial_M^{(2)}\right), \eeqa

\noi such that following relations are satisfied:

\beqa \label{ferm}
\begin{array}{lll}
\left\{\theta^M_{(i)},\theta^N_{(i)}\right\}=0,&
\left[\theta^M_{(1)},\theta^N_{(2)}\right]=0,\\
\left\{\partial_M^{(i)},\theta^N_{(i)}\right\}=\delta_M{}^N,&
\left[\partial_M^{(1)},\theta^N_{(2)}\right]=0,&
\left[\partial_M^{(2)},\theta^N_{(1)}\right]=0,\\
\left\{\partial_M^{(i)},\partial_N^{(i)}\right\}=0,&
\left[\partial_M^{(1)},\partial_M^{(2)}\right]=0.
\end{array}
\eeqa

\noi This means that fermions in the same subspace anticommute,
while fermions in different subspaces commute. The normalisation
in \eqref{green} is chosen to ensure that the relations
\eqref{ferm} reproduce equations \eqref{para1} and \eqref{para}.
Introducing fermionic oscillators $(\eta^M, \partial_{\eta^M})$,
the Green ansatz can be  reformulated as

\beqa \label{green2} \theta^M = \frac{1}{\sqrt{2}}
\left(\eta^M \otimes 1 + 1 \otimes \eta^M\right), \ \
\partial_M = \frac{1}{\sqrt{2}}
\left(\partial_{\eta^M} \otimes 1 + 1 \otimes
\partial_{\eta^M}\right).
\eeqa

\noi It is important to emphasise that, with the representation
\eqref{ferm}, we gain quadratic relations between the
paragrassmann variables and their derivatives. This implies in
particular that

\beqa
\partial_\mu(\theta^\nu)= 2 \delta_\mu{}^\nu,\ \
\partial_\mu(\theta^\nu \theta^\rho)= 2 \delta_\mu{}^\nu\theta^\rho,\ \
\partial_1((\theta^2)^2 \theta^1)=-2 (\theta^2)^2, \   \
\eeqa

\noi
{\it etc.}, holds. Within the Green ansatz, the supercharges then become

\beqa \label{supercharge} V=  \e^\mu_i \partial_\mu^i -(\theta_2
\e_1^\mu + \theta_1 \e_2^\mu) \theta_1. \theta_2 P_\mu. \eeqa

It is obvious that an arbitrary monomial in the paragrassmann
variables can be written as a monomial in its Green component, but
the converse is of course generally wrong. The precise relation
between these two types of polynomials have been studied in detail
in \cite{para}. As a consequence, the set of polynomials in the
Green components is larger than the set of polynomials in the
paragrassmann variables. This opens the possibility to define an
``extended'' superfield which depends on the Green components of
the parafermions $\theta$ instead of $\theta$ itself:

\beqa \label{superPhi-green} \tilde \phi(x^\mu,
\theta_1^M,\theta_2^M). \eeqa

\noi Observe that this implies that the superfield
\eqref{superPhi-green} looks like an $N=2$ superfield. In this
analogy, there are however some differences that should be
observed carefully. Indeed, the variables $\theta_i^M$ are Lorentz
vectors and $\theta_1^M$ commute with $\theta_2^M$. With this
decomposition, we have the identity

\beqa \label{superPhi-green-decompo} \tilde \phi(x^\mu,
\theta_1^M,\theta_2^M)= \sum \limits_{p_1,p_2=0}^4
A_{[p_1,p_2]}{}_{M_1 \cdots M_{p_1}; N_1 \cdots N_{p_2}
}\theta_1^{M_1} \cdots  \theta_1^{N_{p_1}}
 \theta_2^{N_2} \cdots  \theta_2^{N_{p_2}}.
\eeqa

\noi It should be taken into account that, in the decomposition
above, the tensors $A_{[p_1,p_2]}$ are not in irreducible
representations of $GL(1,4)$ (and {\it a forciori} of $SO(1,3)$).
It is however not difficult to decompose the above product. Using
the isomorphism between a $p-$ and an $(4-p)-$form, only a few
products have to be identified. In contrast to the superfield
\eqref{superPhi}, here we find that multiplicities for the
component representations can be greater than one, {\it i.e.}, we
no more obtain multiplicity free reductions.

\medskip
In any physical application, a central object in the construction
of invariant Lagrangians is the covariant derivative, which
commutes with $V$. Indeed, if an operator $D$ such that the
condition $[D,V]=0$ is satisfied can be found, then the latter can
be interpreted as a covariant derivative. This is a consequence of
the Jacobi identity, which in connection with the considered
operator $D$ implies the following relation:

$$
\delta [D,\Phi]= [Q,[D,\Phi]]= [D,[Q,\Phi]]=[D,\delta \Phi].
$$

\noi A routine but cumbersome computation shows that the
construction of such a covariant derivative using \eqref{rep} does
not work. However, if we insist to obtain a covariant derivative,
the obstructions can be surmounted by slightly modifying $V$ in
equation \eqref{rep}. An admissible variation for this purpose is
given if we define the corresponding operator as:

\beqa \label{tercov} V&=&[\epsilon^\mu,\partial_\mu]+
\big([\theta,\epsilon^\mu][\theta_\mu,\theta^\alpha] +
             [\theta,\theta_\mu]  [\e^\mu,\theta^\alpha] +
             [\theta,\theta^\mu] [\theta^\mu,\e^\alpha]\big)P_\alpha,
\nonumber \\
D&=&[\theta,\partial]+[\theta,\theta^\mu][\theta_\mu,\theta^\alpha]
P_\alpha, \eeqa

\noi where $\partial$ is the ``derivative'' associated to the
variable $\theta$. Now, using the relation
$[[\theta^\alpha,\theta^\beta],\theta^\gamma]=0$, we finally get
after some computation to the desired identity $[D,V]=0$. This
fact enables us to define a constraint superfield as a superfield
satisfying the condition

$$
[D,\Phi_c]=0.
$$

\noi Now observe that, since the $y^\mu$ are commuting variables
and the conditions $[D,\theta^\mu]=0$ and $[D,y^\mu]=0$ for
$y^\mu=x^\mu - [\theta,\theta^\nu][\theta_\nu,\theta^\mu]$, the
preceding commutator means that $\Phi_c$ takes a particularly
simple expression

\beqa \label{phic} \Phi_c(y^\mu,\theta^\mu) \eeqa

\noi and does not depend explicitly on the variable $\theta$.
 This
has the interesting consequence concerning the decomposition of
the field \eqref{phic} upon the rules given previously, namely,
that only $GL(1,3)$ tensors have to be considered.
Following this procedure it follows, in particular, that $\lb
V_1,V_2,V_3 \rb. x^\mu $ vanishes identically. This means in particular
that the cubic extension of the Poincar\'e algebra associated to
\eqref{tercov} induces a space-time translation of a null vector.
This point will be commented at the end of this section.

\subsection{Quaternary superfields}

The construction of quaternary superfields goes along the same
lines as the construction of ternary superfields. We would like
however to mention some interesting features. Consider the $D=10$
case. Looking at the  supercharges given by \eqref{paraD=10} or
\eqref{superD=10}, it is not difficult to see that a covariant
derivative commuting with $Q_A$ (or with $Q$) cannot be found.
However, as done already for the ternary case, a slight
modification of equation \eqref{paraD=10} or \eqref{superD=10}
allows us to find operators that can be seen as covariant
derivatives. We illustrate the procedure with paragrassmann
variables. The key step is to introduce

\beqa \label{cov} Q&=&[\epsilon^A,\partial_A] + C_-{}_{AB}
[\theta^A,\theta^B] [\epsilon^C,\theta^D](\Gamma^MC_-)_{CD} P_M -
C_-{}_{AB} [\epsilon^A,\theta^B]
[\theta^C,\theta^D](\Gamma^MC_-)_{CD} P_M,
\nonumber \\
D_A&=&\partial_A -C_-{}_{AB} \theta^B
[\theta^C,\theta^D](\Gamma^MC_-)_{CD} P_M +C_-{}_{BC}
[\epsilon^B,\theta^C] \theta^D (\Gamma^MC_-)_{DA} P_M. \eeqa

\noi Using \eqref{para1}, a direct computation shows that
$[D_A,Q]=0$ holds. Further, since the R.H.S of \eqref{cov}
involves commutators, we obtain that the action of $Q_1 Q_2 Q_3
Q_4$ on the space-time vanishes because of the identity

$$
[Q_1, [Q_2, [Q_2, [Q_4,x^M]]]]=0.
$$

\noi At this point we observe a quite interesting
 property (analogous to what happens in the cubic case) that
arises at once from this consideration: the quartic extension of
the Poincar\'e algebra considered above induces a space-time
translation of a null vector.

Another remarkable consequence of this modification concerns the
algebraic structure of the extension. In contrast to the
previously analyzed ternary case, here a $\mathbb Z_2 \times
\mathbb Z_2 \times\mathbb Z_2 \times\mathbb Z_2$-grading is enough
to ensure the closure of the algebra via
$$
[Q_1,Q_2,Q_3,Q_4] = \sum \limits_{\sigma \in S_4} \epsilon(\sigma)
Q_{\sigma(1)}Q_{\sigma(2)}Q_{\sigma(3)}Q_{\sigma(4)},
$$

\noi where $\epsilon(\sigma)$ denotes the signature of the
permutation $\sigma$. With these specifications, we avoid
completely the use of the complexification of the algebra.\\

Before closing this section, some comments are in order. As we
have seen, if we impose the existence of a covariant derivative
either in the cubic case or the quartic case, we automatically get
that the higher degree extensions considered in this paper
generate naturally a space-time translation of null vectors. Such
a possibility is obviously excluded in usual supersymmetry.
Indeed, recall that when we are studying massless representations
(or massive representations with appropriate central charges, the
so-called BPS-saturated states) half of the generators are
inactive and generate a nilpotent subalgebra. The unitarity of the
representation (absence of ghosts) forces us to represent the
inactive charges by zero (see {\it e.g.} \cite{wb}). However,
since we are considering here cubic and quartic algebras, it is
not obvious at that the argument above remains valid. The
consideration of higher order nilpotent extensions of the
Poincar\'e algebra deserves further investigation. Let us mention
that some kind of ``nilpotent supersymmetry'' in connection with
the previous remark has already been considered, in the context of
pure spinors and BRST symmetry (see \cite{pure} and references
therein).

\section{Conclusions}

We have shown that parafermions  are the relevant variables to
construct an adapted superspace for higher order extensions of the
Poincar\'e algebra (order two parafermions for cubic extensions
and order three parafermions for quartic extensions). In
particular this means that we were able to construct a
differential realisation of the algebras \eqref{3poin} and
\eqref{4poin} leading to an appropriate superspace. Among the
quartic extensions, the $D=10$ case presents some interesting
similarities with algebraic structures considered in supersymmetric
theories \cite{IIA}.

\smallskip

For the classes of extensions considered, we have further analyzed
the possibility of defining a covariant derivative, and have shown
that such a fundamental object exits only if the algebra is
``nilpotent'' in the sense that it implies a space-time
translation of null vector. This phenomenon is therefore deeply
connected with the inner structure of the extension and
superfields, and its range of validity has still to be explored.
Although this point constitutes a fundamental difference when
compared to the requirements of usual supersymmetry, we have to
take into account that the transition to higher order extensions
delivers new possibilities and structural properties, to which the
usual phenomenological interpretations are no longer applicable
automatically. For this reason, it cannot be inferred that the
existence of such translations of null-vectors is intrinsically
devoid of physical meaning.

\smallskip

Having identified the appropriate superfields associated to higher
order extensions of the Poincar\'e algebra, one may wonder whether
the standard techniques will be useful in the construction of
physical models; since the product of superfields is a superfield,
there is in principle no formal difficulty to construct an
interacting theory.

Before closing this paper, let us make a final remark. As we have
seen, parafermions become central for cubic and quartic extensions
of the Poincar\'e algebra. In this context, the question
 whether this procedure can be generalized to higher order extensions
arises at once. More specifically, one can ask whether the fully
symmetric extensions of order $F$ (based upon Lie algebras of
order $F$) also imply the possibility of constructing a
differential realisation based on  order $F-1$  parafermions
operators. More generally, higher order extensions with fully
antisymmetric brackets which   can be seen
 as a special case of the colour algebras introduced
in \cite{color} can be considered. In a straight analogy one may
wonder if parabosons would constitute the relevant variables in
these cases.

To give an argument towards a positive answer to the last
question, consider the algebra $I\mathfrak{so}(1.3) \oplus \left<
W_\mu, \mu=0,\cdots,3\right>$, that is, the  Poincar\'e algebra
together with Lorentz a vector $W_\mu$. Assume furthermore that
the fully antisymmetric  trilinear brackets between the operators
$W$ close upon a space-time translation

\beqa
\label{parb}
[W_\mu,W_\nu,W_\rho] = \epsilon_{\mu \nu \rho \sigma} P^\sigma,
\eeqa

\noi with $[A,B,C]= ABC + BCA + CAB -ACB -BAC -CBA$ and satisfy the identity

$$
[W_\mu,[W_\nu,W_\rho,W_\sigma]]-[[W_\nu,[W_\rho,W_\sigma,W_\mu]]+
[W_\rho,[W_\sigma,W_\mu,W_\nu]]-[W_\sigma,[W_\mu,W_\nu,W_\rho]]=0.
$$
\noi
This  real algebra appears as a special case of the colour algebras introduced
in \cite{color}. Now all the results of section \ref{ternary} can be applied
directly with the following substitutions:  the fully symmetric
brackets $\{\cdots \}$ have to be substituted by the fully antisymmetric
brackets $[\cdots]$ and  commutators $[\theta,\partial], [\theta,\theta]$
{\it etc} by the anticommutators $\{\theta,\partial\}, \{\theta,\theta\}$.
For instance we have

\beqa
\label{parb2}
W=\left\{\e^\mu,\partial_\mu\right\}+ \left\{\theta,\theta^\mu\right\}
\left\{\e^\sigma, \theta_\mu\right\}P_\sigma,
\eeqa

\noi
for the supercharge.
This means that the algebra \eqref{parb}-\eqref{parb2}
 can be realised in terms of the
order two parabosons $\theta, \partial$. However there is two
differences compared to the algebra \eqref{3poin}. Firstly there
is no need to make a kind of ``Jordan-Wigner'' transformation
since $\sum_{\sigma \in S_3} \epsilon(\sigma)W_{\sigma(1)}.
W_{\sigma(2)}.W_{\sigma(3)}.(\theta^{\alpha_1} \cdots
\theta^{\alpha_n})=0$. Secondly, although there is an analogous
theorem for the decomposition of parabosons (see footnote
\ref{paraf}) the corresponding superfield has an infinite number
of degrees of freedom. Furthermore, it as to be noticed that this
differential realisation induces a space-time translation of null
vectors.

\smallskip

In conclusion, let us mention that the construction outlined here
(see \eqref{parb} and \eqref{parb2}), together with the results
obtained in this paper, suggest that parafermions and parabosons
could play some role in the description of higher order
symmetries, but in a different context to its historical
consideration and use.

\section*{Acknowledgments}
During the preparation of this work, one of the authors (RCS) was
financially supported by the research projects MTM2006-09152 of
the M.E.C. and GR58/4120818-920920 of the UCM-BSCH.


\begin{thebibliography}{99}
%
\bibitem{GeN} G. Gentile, Nuovo Cimento {\bf 17} (1940) 493.
%
\bibitem{NE} C. A. Nelson, { J. Phys. } {\bf A37} (2004)
2497.
%
\bibitem{green}
H.~S.~Green, Phys.\ Rev.\  {\bf 90} (1953) 270.
%
\bibitem{para}
Y.~Ohnuki and S.~Kamefuchi, {\it Quantum Field Theory and
Parastatistics} {\it  Tokyo, Japan: Univ. Pr. (1982) 
Berlin, Germany: Springer ( 1982)
 489p}.
%
\bibitem{GrB} O. W. Greenberg, { Phys. Rev. Lett.}
\textbf{13} (1984) 598.
%
\bibitem{GM} O. W. Greenberg  and A. M. L. Messiah, { Phys. Rev.}
\textbf{B136} (1964)  248;
\textbf{B138} (1964)   1155.
%
\bibitem{no-go}
S.~Coleman and J.~Mandula, { Phys. Rev.} {\bf 159} (1967) 1251;
R.~Haag, J.~T.~Lopuszanski and M.~F.~Sohnius, { Nucl. Phys.} {\bf B88}
(1975) 257.
%
%
\bibitem{bl}
 J.~Bagger and N.~Lambert,
  Phys.\ Rev.\   {\bf D75} (2007) 045020
  [arXiv:hep-th/0611108].
%
\bibitem{flie}
M.~Rausch de Traubenberg, M.~J.~Slupinski, J.\ Math.\ Phys.\  {
\bf 41} (2000) 4556 [arXiv:hep-th/9904126];
 M.~Rausch de Traubenberg, M.~J.~Slupinski,
J.\ Math.\ Phys.\  {\bf 43} (2002) 5145 [arXiv:hep-th/0205113].
%
%
\bibitem{gr}
M.~Goze, M.~Rausch de Traubenberg and A.~Tanasa, J. Math. Phys.
{\bf 48} (2007) 093507
  [arXiv:math-ph/0603008].
%
\bibitem{hopf}
M. Rausch de Traubenberg,  J.\ Phys.\ Conf.\ Ser.\  {\bf 128}
(2008) 012060 [arXiv:0710.5368 [math-ph]];
M.~Goze and M. Rausch de Traubenberg,
 J. Math. Phys. {\bf 50} (2009) 063508
  [arXiv:0809.4212 [math-ph]].
%
\bibitem{color}
R.~Campoamor-Stursberg and M.~Rausch de Traubenberg,
J. of Generalized Lie Theory and Appl. {\bf 3} (2009) 113
  [arXiv:0811.3076 [math-ph]].
%
\bibitem{cubic1}
N.~Mohammedi, G.~Moultaka and M.~Rausch de Traubenberg, Int.\ J.\ Mod.\
Phys.\ {\bf A19}  (2004) 5585 [arXiv:hep-th/0305172].
%
\bibitem{cubic2}
G.~Moultaka, M.~Rausch de Traubenberg and A.~Tanasa,  Int.\ J.\ Mod.\
Phys.\  {\bf A20} (2005) 5779 [arXiv:hep-th/0411198].
%
\bibitem{p-form}
G.~Moultaka, M.~Rausch de Traubenberg and A.~Tanasa, {Proceedings
of the XIth International Conference Symmetry Methods in Physics,
Prague 21-24 June 2004} [arXiv:hep-th/0407168].
%
\bibitem{super}
C.~Ahn, D.~Bernard and A.~ Leclair,
 Nucl. Phys. {\bf B346} (1990) 409;
%
S.~ Durand,
 Mod. Phys. Lett {\bf A8}  (1993) 2323 [hep-th/9305130];
%
N.~ Fleury and M. Rausch de Traubenberg,
 Mod. Phys. Lett. {\bf A11} (1996)
899 [hep-th/9510108];
%
A.~Perez, M. Rausch de Traubenberg and P. Simon,
 Nucl. Phys. {\bf B482} (1996) 325 [hep-th/9603149];
%
M. Rausch de Traubenberg and P. Simon, Nucl. Phys. {\bf B517} (1998) 485 [hep-th/9606188].
%
\bibitem{az}

J.~A.~ de Azc\'arraga and A.~J.~ Macfarlane,
J. Math. Phys. {\bf 37} (1996) 1115 [hep-th/9506177];
%
 R.~S.~Dunne, A.~J.~Macfarlane, J.~A.~de Azcarraga and J.~C.~Perez Bueno,
   Phys.\ Lett.\   {\bf B 387} (1996) 294
  [arXiv:hep-th/9607220];
%
R.~S.~Dunne, A.~J.~Macfarlane, J.~A.~de Azc\'arraga and
J.~C.~P\'erez Bueno,  {  Int. J.  Mod. Phys. } {\bf A12} (1997) 3275
[hep-th/9610087].
%
\bibitem{roby} N. Roby,  Bull. Sc. Math. \textbf{94} (1970) 49.
%
\bibitem{tern}
I. Bars   and  M. G\"un\"aydin, {  J.\ Math.\ Phys.\ }  {\bf
20} (1979)  1977;
I. Bars   and M.  G\"un\"aydin, {   Phys.\ Rev.\   } {\bf D22}
(1980)  1403;
%
 T.~D.~Palev,
J.\ Math.\ Phys.\  {\bf 23} (1982) 1100;
%
N.~I.~Stoilova and J.~Van der Jeugt, Journal of Physics:
Conference Series {\bf 128} (2008) 012061
  [arXiv:math-ph/0611085].
%
\bibitem{parag}
%
 J.~Beckers and N.~Debergh,
  Int.\ J.\ Mod.\ Phys.\   {\bf A8}  (1993) 5041;
%
A.~G.~Nikitin and V.~V.~Tretynyk, { J.\ Phys.} {\bf A28}
(1995) 1655;
%
J. Niederle and A.~G.~Nikitin, { J.\ Phys.} {\bf A32}  (1999) 5141.
%
\bibitem{ply}
M.~Plyushchay,    Int.\ J.\ Mod.\ Phys.\   {\bf A15} (2000) 3679
  [arXiv:hep-th/9903130];
 S.~Klishevich and M.~Plyushchay,
  Mod.\ Phys.\ Lett.\   {\bf A14} (1999) 2739
  [arXiv:hep-th/9905149];
F.~Correa, V.~Jakubsky, L.~M.~Nieto and M.~S.~Plyushchay,
  Phys.\ Rev.\ Lett.\  {\bf 101} (2008) 030403
  [arXiv:0801.1671 [hep-th]];
F.~Correa, V.~Jakubsky and M.~S.~Plyushchay,
  J.\ Phys.\   {\bf A41} (2008) 485303 
  [arXiv:0806.1614 [hep-th]].

%
\bibitem{noether}
M.~Rausch de Traubenberg, Pr. Inst. Mat. Nats. Akad. Nauk Ukr. Mat. Zastos.,
50, Part 1, 2, 3, Natsional. Akad. Nauk Ukraïni, \~Inst. Mat.,
Kiev, 2004, pp. 578 [arXiv:hep-th/0312066].
%
\bibitem{cours}
 M.~Rausch de Traubenberg,
  J. Phys. Conf. Series {\bf 175} (2009) 012003 [arXiv:0811.1465 [hep-th]].
\bibitem{coloralg}
V.~Rittenberg and D.~Wyler,
  J.\ Math.\ Phys.\  {\bf 19}  (1978) 2193;
%
V.~Rittenberg and D.~Wyler,
  Nucl.\ Phys.\  {\bf B139} (1978) 189;
%
M.~Scheunert,
  J.\ Math.\ Phys.\  {\bf 20}  (1979) 712;
%
H.~S.~Green and P.~D.~Jarvis,
   J.\ Math.\ Phys.\  {\bf 24} (1983) 1681;
%
J.~Lukierski and V.~Rittenberg,
  Phys.\ Rev.\  {\bf D18} (1978) 385.
%
\bibitem{kern}
 R.~Kerner,
  J.\ Math.\ Phys.\  {\bf 33} (1992) 403,
%
 V.~Abramov, R.~Kerner and B.~Le Roy,
 J.\ Math.\ Phys.\  {\bf 38} (1997) 1650
 [arXiv:hep-th/9607143].
%
\bibitem{IIA}
 I.~C.~G.~Campbell and P.~C.~West,
  Nucl.\ Phys.\   {\bf B243} (1984) 112;
 M.~Huq and M.~A.~Namazie,
  Class.\ Quant.\ Grav.\  {\bf 2} (1985) 293
  [Erratum-ibid.\  {\bf 2} (1985) 597];
 F.~Giani and M.~Pernici,
  Phys.\ Rev.\   {\bf D30} (1984) 325.
%
\bibitem{young}
M. Hamermesh, {\it Group theory and its application to physical
problems} (Addison-Wesley, 1962); W. Fulton and J. Harris, {\it
Representation theory: a first course} (Springer, 2004).
%
\bibitem{theo1}
Y.~Ohnuki and S. Kamefuchi, { Phys. Rev.} {\bf 170} (1968)
1279.
\bibitem{theo2}
S.~Kamefuchi and Y. Takahashi, { Prog. Theor. Phys.} (Kyoto)
Suppl. {\bf 37 } \& {\bf 38} (1966) 244.
%
\bibitem{cliff}
 M. Rausch de Traubenberg,   to appear in Adv. Appl. Cliff. Algebra 
 [arXiv:hep-th/0506011].
%
\bibitem{wb}
J. Wess, J. Bagger, {\it ``Supersymmetry and
Supergravity''} (Princeton University Press, 1983).
%
\bibitem{pure}
M. Cederwall, arXiv:0906.5490 [hep-th].



\end{thebibliography}
\end{document}